\documentclass[conference]{IEEEtran}
\IEEEoverridecommandlockouts
\usepackage{cite}
\usepackage{amsmath,amssymb,amsfonts}
\usepackage{algorithmic}
\usepackage{graphicx}
\usepackage{textcomp}
\usepackage{booktabs} 
\usepackage{pgfplots}
\usepackage{float}
\usepackage{multirow}
\usepackage{siunitx}

\usepackage[top=0.75in, bottom=1in, left=1in, right=1in]{geometry}
\definecolor{blue}{rgb}{0.0, 0, 0}

\IEEEoverridecommandlockouts

\pgfplotsset{compat=1.18}

\usepackage{xcolor}
\def\BibTeX{{\rm B\kern-.05em{\sc i\kern-.025em b}\kern-.08em
    T\kern-.1667em\lower.7ex\hbox{E}\kern-.125emX}}
\begin{document}

\title{A Framework Leveraging Large Language Models for Autonomous UAV Control in Flying Networks
\thanks{This work is financed by National Funds through the FCT -- Fundação para a Ciência e a Tecnologia, I.P. (Portuguese Foundation for Science and Technology) within the project FALCON, with reference 2023.15645.PEX (https://doi.org/10.54499/2023.15645.PEX).}
}

\author{\IEEEauthorblockN{Diana Nunes, Ricardo Amorim, Pedro Ribeiro, André Coelho, Rui Campos}
\IEEEauthorblockA{INESC TEC and Faculdade de Engenharia, Universidade do Porto, Portugal\\
\{diana.m.nunes, ricardo.a.amorim, pedro.m.ribeiro, andre.f.coelho, rui.l.campos\}@inesctec.pt}
}

\maketitle

\begin{abstract}
This paper proposes FLUC, a modular framework that integrates open-source Large Language Models (LLMs) with Unmanned Aerial Vehicle (UAV) autopilot systems to enable autonomous control in Flying Networks (FNs). FLUC translates high-level natural language commands into executable UAV mission code, bridging the gap between operator intent and UAV behaviour.

FLUC is evaluated using three open-source LLMs -- Qwen 2.5, Gemma 2, and LLaMA 3.2 -- across scenarios involving code generation and mission planning. Results show that Qwen 2.5 excels in multi-step reasoning, Gemma 2 balances accuracy and latency, and LLaMA 3.2 offers faster responses with lower logical coherence. A case study on energy-aware UAV positioning confirms FLUC’s ability to interpret structured prompts and autonomously execute domain-specific logic, showing its effectiveness in real-time, mission-driven control. 
\end{abstract}


\section{Introduction}

\begin{figure}
    \centering
    \includegraphics[width=0.94\linewidth]{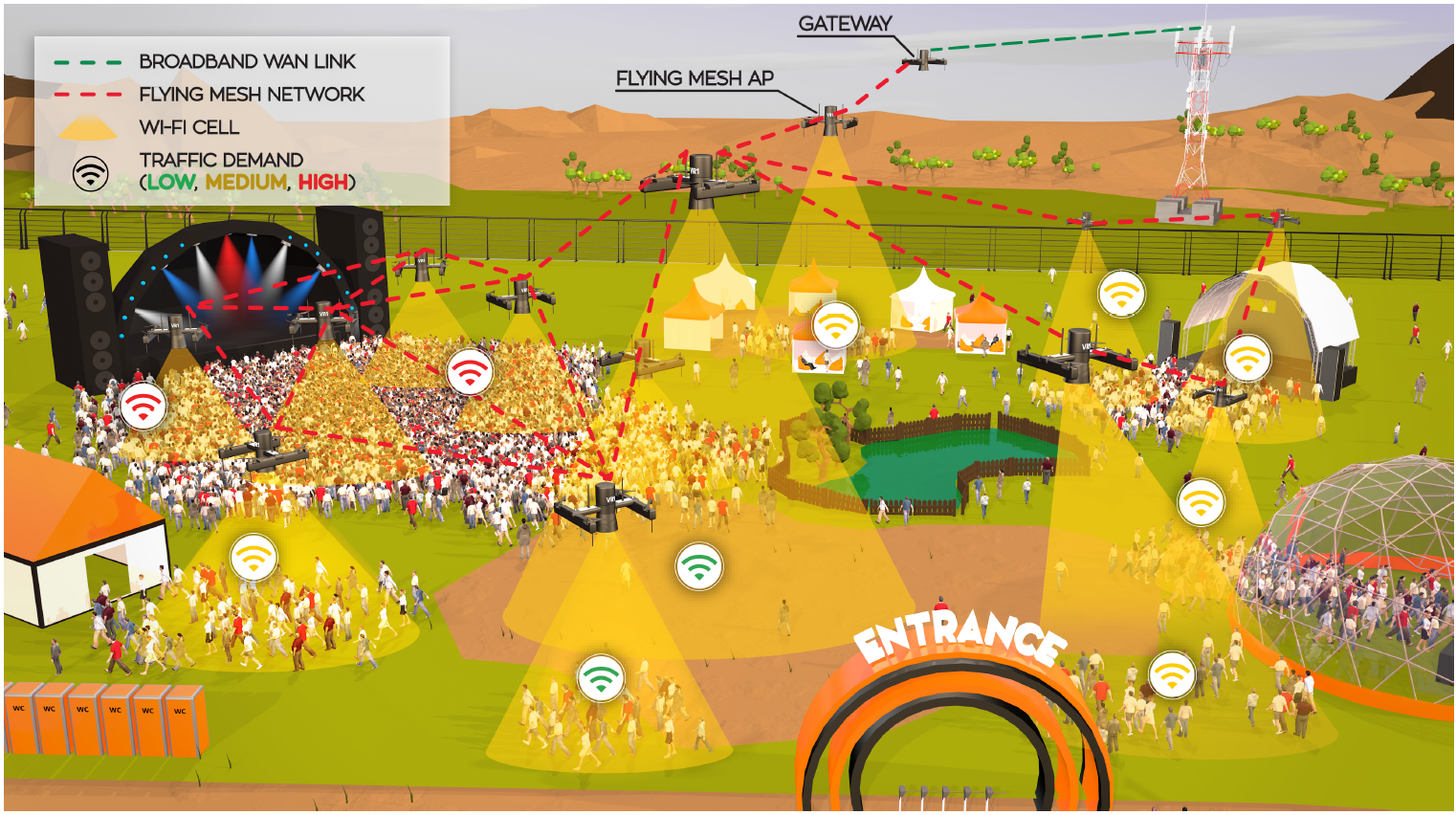}
    \caption{Flying network providing on-demand broadband wireless connectivity in a large-scale event.}
    \label{fig:example}
\end{figure}

The demand for adaptable and reliable wireless communications systems has led to the adoption of Flying Networks (FNs), where Unmanned Aerial Vehicles (UAVs) act as airborne communications nodes. FNs provide on-demand network coverage in scenarios where terrestrial infrastructure is infeasible or insufficient, such as disaster response, large-scale events, and remote rural areas (see Fig.~\ref{fig:example}). Due to their mobility and rapid deployment, FNs are increasingly used to support scalable, mission-critical communications services.

Recent advances in artificial intelligence, especially Large Language Models (LLMs), have demonstrated significant capabilities in domains such as autonomous driving, logistics, and automated reasoning~\cite{Sun2024, Rong2024, Vemprala2024}. However, their application in UAV-based communications systems remains limited~\cite{Javaid2024_2, Sun2024_2}, presenting a research opportunity to explore language-driven autonomy in airborne networking. In particular, LLMs’ ability to convert natural language into structured control logic enables intuitive interaction with autonomous systems, reducing the need for domain-specific programming and human effort.

This paper presents FLUC, a modular framework that integrates open-source LLMs with the ArduPilot autopilot stack~\cite{ardupilot} to enable autonomous UAV control in FNs. FLUC interprets high-level natural language instructions and converts them into executable mission code. The framework supports both simulation and real-world deployment and operates offline via the Ollama runtime. 

The main contributions of this work are as follows:
\begin{itemize}
    \item \textbf{FLUC}, a modular LLM-based framework that enables natural language UAV control, lowering barriers of entry and advancing autonomous operations in FNs;
    \item \textbf{A comparative evaluation of Qwen 2.5 Coder, Gemma 2, and LLaMA 3.2}, assessing their effectiveness in generating structured UAV control logic for communications-driven tasks.
\end{itemize}

To the best of our knowledge, FLUC is the first implementation of LLM-based UAV control using locally deployed models. This work demonstrates the feasibility of interpreting high-level intent 
 within FNs using LLMs and paves the way for future research on autonomous, language-driven control of wireless networks.

The rest of the paper is structured as follows. Section~\ref{RW} reviews related work on LLM-based control and UAV autonomy. Section~\ref{PF} details the FLUC architecture. Section~\ref{LE} outlines the evaluation setup and results. Section~\ref{CS} presents a case study on UAV positioning. Finally, Section~\ref{C} concludes with the main findings and future research directions.

\section{Related Work}\label{RW}

The integration of LLMs into robotic systems has attracted growing interest, particularly for enabling natural language-based control and intuitive human-machine interaction. Vemprala et al.~\cite{Vemprala2024} explored ChatGPT’s capabilities on robotic platforms, identifying key functions such as logical reasoning, code synthesis, and dialogue grounding. Using the AirSim simulator, they demonstrated LLM-based control of aerial and ground robots through high-level text instructions, showing the feasibility of natural language-driven control in simulated environments.

In the UAV domain, Javaid et al.~\cite{Javaid2024} investigated LLMs for autonomous flight control, onboard perception, and decision support. Other studies explored direct integration of LLMs with UAV systems. Tazir et al.~\cite{Tazir2023} linked ChatGPT with PX4 and Gazebo to implement natural language-based UAV control, while Piggott et al.~\cite{Piggott2023} introduced Net-GPT, an LLM-powered offensive chatbot that generates context-aware network packets for man-in-the-middle attacks between UAVs and ground control stations. While these efforts validate language-based interaction, they remain limited to simulations and do not support real-time mission execution.

In the context of FNs, recent research has applied AI and LLMs to network optimization and multi-agent coordination~\cite{Javaid2024_2, Sun2024_2}. However, most studies focus on abstract coordination or centralized planning, without addressing real-world LLM deployment in UAV systems. The absence of practical implementations limits these efforts to theoretical exploration.

More recently, the LLM ecosystem has expanded to include locally executable, open-source models -- offering alternatives to proprietary cloud-based solutions. While commercial models such as GPT and Gemini perform well, their dependence on cloud APIs introduces latency, privacy, and connectivity issues -- constraints that may be unsuitable for edge-deployed UAVs. In contrast, lightweight models like LLaMA 3.2~\cite{llama3.2}, Gemma 2~\cite{gemma2}, and Qwen 2.5 Coder~\cite{qwen2.5} support offline execution with competitive performance. LLaMA 3.2 provides low-latency output for resource-limited devices; Gemma 2 balances reasoning and speed; and Qwen 2.5 Coder excels in structured code generation. These features make them suitable for onboard UAV autonomy.

In terms of flight optimization, several studies address Quality of Service (QoS) and energy-aware UAV trajectories~\cite{Ribeiro2024, Rodrigues2022}. However, these assume manual control logic and do not explore autonomous execution of planned missions.

To our knowledge, prior work lacks a practical implementation of UAV control driven by locally deployed LLMs.

\section{Proposed Framework: FLUC} \label{PF}

We propose FLUC, a modular framework that enables edge devices to interpret and execute natural language instructions autonomously, without reliance on cloud services. By leveraging locally deployed LLMs, FLUC converts high-level user intent into structured control logic, allowing non-expert users to deploy missions across both simulation and real-world environments. It serves as a practical interface between natural language and UAV operations.

\subsection{System Architecture}

FLUC adopts a modular architecture designed for extensibility across UAV platforms and simulation environments. It is compatible with MAVLink~\cite{mavlink}, a standard drone control protocol, and consists of two main components: Ollama runtime~\cite{ollama}, a lightweight inference engine for offline LLM execution with low-latency response and full data privacy; and ArduPilot~\cite{ardupilot}, an open-source autopilot system interfaced via MAVProxy~\cite{mavproxy}, which serves as the control backend for both Software-In-The-Loop (SITL)~\cite{ardupilot_sitl} simulations and real UAV deployments.

\begin{figure*}
    \centering
    \includegraphics[width=0.95\linewidth]{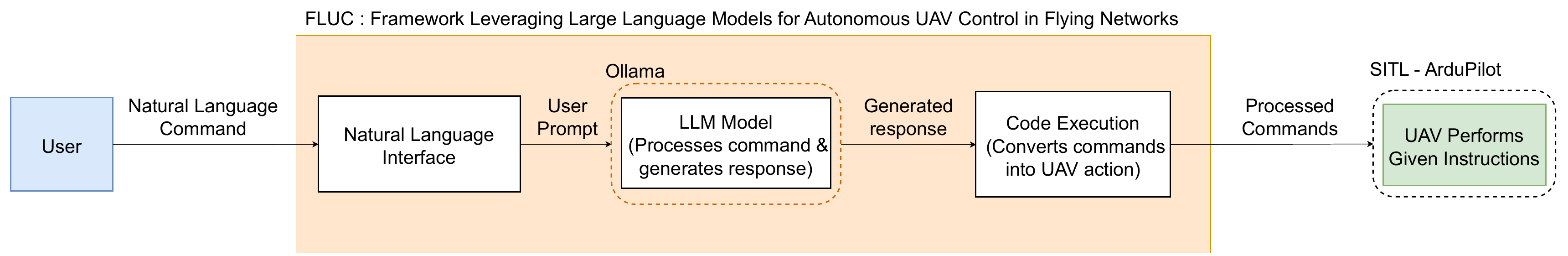}
    \caption{FLUC system architecture: from user input to mission execution via local LLM processing and UAV control logic.}
    \label{fig:framework}
\end{figure*}

Users interact with FLUC via a terminal interface, issuing natural language prompts that define mission objectives. These prompts are parsed by the LLM in Ollama, which generates executable Python code. A predefined function library -- exposed through prompt engineering -- includes primitives such as \texttt{go\_to\_real\_world\_coords()}, \texttt{move\_relative()}, and \texttt{set\_return()}. These serve as semantic building blocks for UAV mission logic. The generated code is passed to a mission handler module that interfaces with ArduPilot through MAVProxy. Missions are executed either in SITL or on physical UAV hardware. The overall architecture is shown in Fig.~\ref{fig:framework}.

FLUC supports seamless deployment in MAVLink-compliant environments, including Hardware-In-The-Loop (HITL) testing and live UAV missions.

\subsection{Functional Capabilities}

FLUC enables natural language control, allowing operators to issue mission commands without programming expertise. Local LLM execution ensures low-latency responses and eliminates dependency on cloud services. The system translates operator intent into executable UAV logic via a modular function library. Its architecture also supports interchangeable LLM backends, integration of optimization modules, and extension with new control routines, making it scalable and adaptable for diverse UAV applications.

\subsection{Functional Validation}

To validate FLUC’s operability, a baseline test was conducted using a directional motion task based on the prompt: \textit{``Fly in a straight diagonal line for 200 meters''}.

\begin{figure}
\centering
\includegraphics[width=0.94\linewidth]{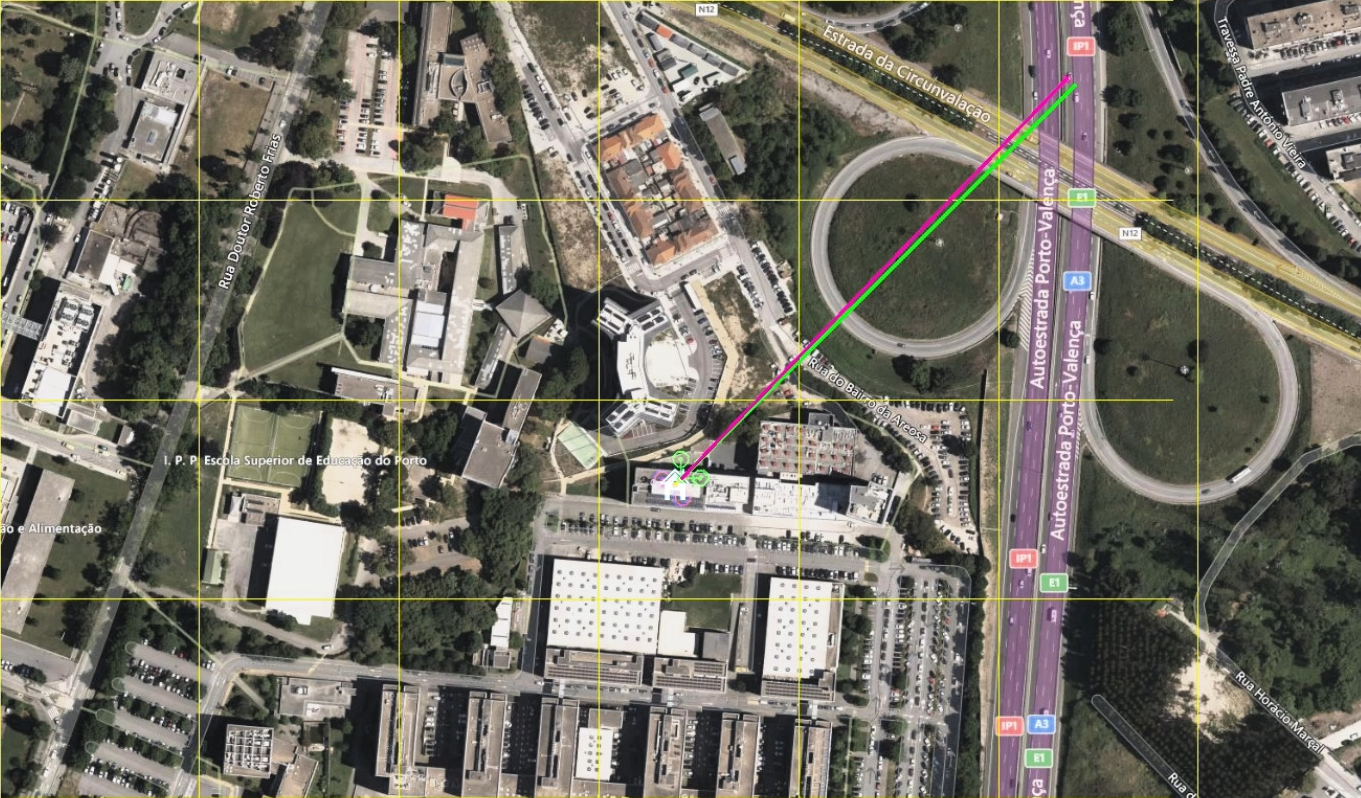}
\caption{UAV trajectory generated in response to a natural language command, validating the FLUC framework.}

\label{fig:trajectory}
\end{figure}

The selected LLM, Qwen 2.5 Coder, running locally via Ollama, generated the corresponding Python code, which was executed in the ArduPilot SITL environment through MAVProxy. The UAV completed the expected diagonal trajectory, confirming accurate translation of natural language into UAV behavior. The resulting flight path is shown in Fig.~\ref{fig:trajectory}.

This test confirms FLUC’s core functionality and establishes a foundation for more complex mission execution and domain-specific extensions.

\subsection{Application Scenarios}

FLUC is suited for FNs, where UAVs act as mobile base stations, relays, or access points requiring adaptive positioning. Natural language-driven control improves deployability in time-sensitive or infrastructure-constrained scenarios.

Beyond wireless communications, FLUC can support applications in surveillance, agricultural mapping, disaster monitoring, and environmental sensing. In each case, the natural language interface reduces technical barriers and enhances mission clarity for operators.

\section{LLM Evaluation} \label{LE}

This section presents a comparative evaluation of three open-source LLMs integrated within the FLUC framework. The objective is to assess the suitability of each LLM for structured mission-oriented UAV tasks with possibly real-time constraints. The evaluation focuses on four metrics -- token usage, execution time, prompt efficiency, and output quality -- across a set of test scenarios that consider realistic practical deployment conditions.

\subsection{LLM Selection}

Model selection was limited by compatibility with the Ollama runtime, which supports low-latency, offline execution -- an important requirement for edge-deployed airborne systems. Three open-source LLMs were selected for evaluation, all with default parameters employed by the Ollama implementation. The first, LLaMA 3.2 (3B), is a compact 3-billion parameter model focused on responsiveness and efficient inference. The second, Gemma 2 (9B), is a general-purpose decoder-only model with 9 billion parameters. Although not specialized for code synthesis, it offers solid instruction-following capabilities and serves as a representative baseline for benchmarking. The third model, Qwen 2.5 Coder (14B), is explicitly optimized for structured programming tasks. Its inclusion enables assessment of the benefits of domain-specific pretraining in generating UAV control logic and supporting structured reasoning.  

\subsection{Evaluation Scenarios}

To evaluate each model's adaptability and ability to synthesize UAV control logic, five mission scenarios were defined. These scenarios increase in complexity, progressing from basic navigation tasks to advanced reasoning challenges involving spatial awareness and dynamic path planning. The first scenario, coordinates-based navigation, used a prompt such as \textit{``Go to 41.1783107 -8.591609 17''}, requiring the model to convert raw GPS input into executable UAV commands. The second scenario tested semantic geolocation using the prompt \textit{``Go to FEUP''}, which involved resolving a named location via OpenStreetMap~\cite{openstreetmap} and generating appropriate control logic. The third scenario focused on relative motion, with instructions such as \textit{``Fly in a straight diagonal line for 200 meters''}, to explore flight coordinates generation. In the fourth scenario, the models were asked to perform path optimization, synthesizing the most efficient paths through a series of five waypoints with 3D coordinates, thereby testing spatial reasoning. Finally, the fifth scenario introduced obstacle avoidance, requiring the models to apply conditional logic to adapt the UAV’s route in response to environmental constraints.


\subsection{Hardware Specification}

All experiments were conducted on a system running \texttt{Ubuntu 24.04.1 LTS (x86\_64)} with an AMD Ryzen 7 5800HS processor (8 cores / 16 threads at 4.46 GHz), 16 GB RAM, and a dedicated NVIDIA GeForce GTX 1650 Mobile (Max-Q) GPU. This setup represents a reproducible and moderately resource-limited computational environment, reflecting typical conditions for UAV ground control in both research and field-testing scenarios.

\subsection{Token and Time Efficiency}

To assess computational efficiency, each of the five scenarios was executed ten times per LLM. The evaluation focused on two primary metrics: the number of tokens generated per task and the average response time required to synthesize mission code. Table~\ref{tab:tokens} reports the mean token usage per scenario, while Table~\ref{tab:sec} presents the corresponding execution times, both expressed with 95\% confidence intervals.

\begin{table}
\centering
\caption{Average number of tokens used per scenario and model, with 95\% confidence intervals.}
\setlength{\tabcolsep}{4pt}
\begin{tabular}{lccc}
\toprule
\textbf{Scenario} & \textbf{LLaMA 3.2} & \textbf{Gemma 2} & \textbf{Qwen 2.5} \\
\midrule
1: Coordinates & 148.5 $\pm$ 20.7 & 45.8 $\pm$ 0.3 & 90.6 $\pm$ 39.7 \\
2: Location    & 275.9 $\pm$ 57.2 & 33.7 $\pm$ 3.9 & 140.6 $\pm$ 39.8 \\
3: Diagonal    & 108.0 $\pm$ 18.0 & 22.4 $\pm$ 0.5 & 55.8 $\pm$ 24.8 \\
4: Waypoints   & 633.4 $\pm$ 89.4 & 339.0 $\pm$ 55.9 & 582.1 $\pm$ 105.8 \\
5: Obstacle    & 411.6 $\pm$ 75.8 & 97.0 $\pm$ 22.7 & 225.6 $\pm$ 72.7 \\
\bottomrule
\end{tabular}
\label{tab:tokens}
\end{table}

\begin{table}
\centering
\caption{Average execution time (in seconds) per scenario and model, with 95\% confidence intervals.}
\setlength{\tabcolsep}{4pt}
\begin{tabular}{lccc}
\toprule
\textbf{Scenario} & \textbf{LLaMA 3.2} & \textbf{Gemma 2} & \textbf{Qwen 2.5} \\
\midrule
1: Coordinates & \textbf{12.4 $\pm$ 1.8} & 13.4 $\pm$ 0.4 & 32.9 $\pm$ 13.2 \\
2: Location    & 25.1 $\pm$ 5.6 & \textbf{10.5 $\pm$ 0.9} & 49.7 $\pm$ 13.7 \\
3: Diagonal    &  9.0 $\pm$ 1.4 &  \textbf{8.0 $\pm$ 0.9} & 21.7 $\pm$ 8.7 \\
4: Waypoints   & \textbf{68.2 $\pm$ 11.4} & 110.5 $\pm$ 21.4 & 236.4 $\pm$ 40.8 \\
5: Obstacle    & 41.1 $\pm$ 9.0 & \textbf{29.0 $\pm$ 5.7} & 83.0 $\pm$ 25.8 \\
\bottomrule
\end{tabular}
\label{tab:sec}
\end{table}

Among the three models, LLaMA 3.2 consistently generated the highest number of tokens but also offered the fastest execution times. Gemma 2 presented the lowest token usage and stable execution times across all scenarios. In contrast, Qwen 2.5 Coder produced more verbose and context-rich outputs -- particularly in complex tasks -- at the cost of longer inference times. 

Waypoint-based path optimization was the most time-consuming task. Due to the relatively high execution time, this approach may be better suited for pre-planned, non-time-sensitive missions.

\subsection{Prompt Efficiency}

Prompt efficiency was evaluated as the average number of user interactions required to generate an executable response, as summarized in Table~\ref{tab:prompt_counts}. Each interaction reflects a user-issued prompt aimed at correcting an incomplete, incorrect, or non-executable response. This procedure was repeated until a valid output was obtained, regardless of whether the final output successfully fulfilled the intended task.

\begin{table}
\centering
\caption{Average number of prompts required per scenario and model, with 95\% confidence intervals.}
\setlength{\tabcolsep}{4pt}
\begin{tabular}{lccc}
\toprule
\textbf{Scenario} & \textbf{LLaMA 3.2} & \textbf{Gemma 2} & \textbf{Qwen 2.5} \\
\midrule
1: Coordinates & 2.2 $\pm$ 0.3 & \textbf{1.0 $\pm$ 0.0} & \textbf{1.0 $\pm$ 0.0} \\
2: Location    & 2.2 $\pm$ 0.5 & \textbf{1.0 $\pm$ 0.0} & \textbf{1.0 $\pm$ 0.0} \\
3: Diagonal    & 1.8 $\pm$ 0.6 & \textbf{1.0 $\pm$ 0.0} & \textbf{1.0 $\pm$ 0.0} \\
4: Waypoints   & 2.0 $\pm$ 0.5 & 2.0 $\pm$ 1.1         & \textbf{1.8 $\pm$ 0.2} \\
5: Obstacle    & 2.0 $\pm$ 0.3 & 1.2 $\pm$ 0.3         & \textbf{1.1 $\pm$ 0.2} \\
\midrule
\textbf{Average} & 2.0 $\pm$ 0.2 & 1.2 $\pm$ 0.2 & \textbf{1.2 $\pm$ 0.1} \\
\bottomrule
\end{tabular}
\label{tab:prompt_counts}
\end{table}

As shown in Table~\ref{tab:prompt_counts}, both Gemma 2 and Qwen 2.5 required fewer prompt iterations on average compared to LLaMA 3.2, indicating better convergence and lower interaction overhead. This discrepancy became more pronounced in complex scenarios, where LLaMA 3.2 often required nearly twice as many user interventions to produce an executable output.

\subsection{Output Quality}

In addition to quantitative metrics, output quality was assessed through qualitative classification into three categories: \textit{Successful}, \textit{Partially Correct}, and \textit{Unsuccessful}. \textit{Successful} outputs performed the intended action correctly on the first attempt, without needing any adjustments. \textit{Partially Correct} outputs were syntactically valid but presented minor errors, omissions, or misinterpretations that affected full task execution. \textit{Unsuccessful} outputs were either invalid, unrelated to the prompt, or failed to produce runnable mission code.

\begin{table}
\centering
\caption{Output classification (Successful / Partial / Unsuccessful) per model and scenario.}
\setlength{\tabcolsep}{4pt}
\begin{tabular}{lccc}
\toprule
\textbf{Scenario} & \textbf{LLaMA 3.2} & \textbf{Gemma 2} & \textbf{Qwen 2.5} \\
\midrule
1: Coordinates & 4 / 3 / 3 & \textbf{10 / 0 / 0} & 5 / 4 / 1 \\
2: Location    & 2 / 1 / 7 & \textbf{9 / 1 / 0}  & 3 / 1 / 6 \\
3: Diagonal    & 2 / 3 / 5 & \textbf{10 / 0 / 0} & 4 / 0 / 6 \\
4: Waypoints   & 0 / 1 / 9 & 0 / 8 / 2           & \textbf{5 / 5 / 0} \\
5: Obstacle    & 0 / 3 / 7 & 0 / 4 / 6           & \textbf{4 / 4 / 2} \\
\bottomrule
\end{tabular}
\label{tab:success}
\end{table}

As shown in Table~\ref{tab:success}, Qwen 2.5 presented the most consistent performance, particularly excelling in complex tasks such as waypoint optimization and obstacle avoidance. Gemma 2 performed reliably in simpler tasks but its success rate declined as task complexity increased. In contrast, LLaMA 3.2 frequently failed to produce executable outputs in structured scenarios, often requiring multiple attempts and exhibiting a higher rate of incomplete or incorrect responses.

\subsection{Discussion} \label{D}

The results reveal clear trade-offs between model size, computational efficiency, and reliability in UAV mission code generation. While all three models successfully handled basic single-step prompts, their performance diverged significantly as task complexity increased.

Qwen 2.5, a 14B parameter model optimized for code synthesis, demonstrated the most robust performance, consistently generating correct outputs with minimal user intervention. Its strength lies in handling structured reasoning and function composition, making it well suited for mission-critical deployments where output accuracy is prioritized over latency. Gemma 2 performed reliably in scenarios involving predefined logic and structured function calls, but showed limitations in handling open-ended reasoning and novel task synthesis -- reflective of its general-purpose design. While efficient and responsive, its performance decreased under complex conditions. In contrast, LLaMA 3.2, despite offering the fastest response times due to its smaller 3B architecture, frequently failed in structured tasks. Its outputs were often incomplete or improperly formatted, particularly in multi-step reasoning scenarios, leading to lower overall task success.

These findings highlight the importance of aligning model capabilities with deployment requirements. Lightweight models like LLaMA 3.2 may be suitable for fast interaction in low-risk settings, whereas code-specialized models such as Qwen 2.5 are better suited for autonomous, logic-intensive UAV operations. Furthermore, it is important to note the critical role of prompt engineering and output validation. The performance of the models can be further improved through targeted improvements in interaction design.

\section{Case Study: Applying the SUPPLY Algorithm} \label{CS}

To complement the general evaluation, this section presents a case study that demonstrates FLUC’s capability to interface with a state-of-the-art UAV positioning algorithm. The objective is to assess whether an LLM can accurately interpret a structured natural language prompt and invoke a predefined, domain-specific function -- in this case, the state-of-the-art SUPPLY algorithm~\cite{Ribeiro2024}. The focus lies on function comprehension, parameter construction, and end-to-end invocation.

\subsection{SUPPLY Algorithm Overview}

SUPPLY is an energy-aware UAV positioning algorithm developed for FNs. It computes an optimized trajectory that minimizes propulsion energy while ensuring QoS for Ground Users (GUs). The algorithm’s output is a sequence of 3D waypoints that simultaneously satisfy energy-efficiency and communication constraints.

\subsection{Scenario and Prompt Description}

In this case study, each LLM was tasked with initiating a UAV mission involving two GUs, each located at predefined coordinates with specified traffic demands (in Mbit/s). The task was described via a natural language prompt:

\begin{quote}
\textit{"Upload and start the mission for a UAV with 2 GUs, whose x are [25,50], y are [50,50], z are [0,0], and traffic [200,200]."}
\end{quote}

The LLM was expected to extract relevant parameters from the prompt and construct a valid function call to invoke the SUPPLY algorithm. This function was predefined within FLUC and made accessible to the LLM via the initialization prompt.

\subsection{Execution Results}

Table~\ref{tab:supply_success} summarizes the output qualitative classification across ten executions for each model. Both Qwen 2.5 and Gemma 2 achieved full success rates, consistently generating correct and executable function calls. In contrast, LLaMA 3.2 was not able to produce a valid execution in any attempt, with outputs classified as either partially correct or unusable due to missing parameters or syntactic errors.

\begin{table}
\centering
\caption{Output classification in the SUPPLY case study (10 attempts per model).}
\setlength{\tabcolsep}{6pt}
\begin{tabular}{lccc}
\toprule
\textbf{Model} & \textbf{Successful} & \textbf{Partial} & \textbf{Unsuccessful} \\
\midrule
LLaMA 3.2     & 0  & 4 & 6 \\
Gemma 2       & \textbf{10} & 0 & 0 \\
Qwen 2.5      & \textbf{10} & 0 & 0 \\
\bottomrule
\end{tabular}
\label{tab:supply_success}
\end{table}

Table~\ref{tab:supply_efficiency} presents the average token usage and execution time per model for the SUPPLY case study. Qwen 2.5 led to the highest token count and latency, consistent with its verbose and detail-oriented code generation style. While LLaMA 3.2 achieved the lowest execution times, its outputs were of lower quality and frequently omitted required syntax elements (e.g., missing code block delimiters such as \texttt{```python}), requiring repeated prompting, which contributed to higher overall token usage.

\begin{table}
\centering
\caption{Average token usage and execution time in the SUPPLY case study, with 95\% confidence intervals}
\setlength{\tabcolsep}{6pt}
\begin{tabular}{lcc}
\toprule
\textbf{Model} & \textbf{Tokens} & \textbf{Time (s)} \\
\midrule
LLaMA 3.2     & 113.9 $\pm$ 10.7 & 9.7 $\pm$ 0.9 \\
Gemma 2       & 67.9  $\pm$ 3.1  & 20.2 $\pm$ 1.0 \\
Qwen 2.5      & 91.8 $\pm$ 29.0  & 35.5 $\pm$ 10.6 \\
\bottomrule
\end{tabular}
\label{tab:supply_efficiency}
\end{table}

These findings align with the broader evaluation trends. Code-specialized models such as Qwen 2.5 demonstrated strong reliability in executing structured logic, particularly in function-driven tasks. General-purpose models like Gemma 2 performed well when prompts were well-defined and scope-limited, but showed reduced robustness when handling ambiguous or generative instructions. Lightweight models such as LLaMA 3.2, while offering rapid responses, were less dependable in syntax-sensitive operations that demanded precise formatting and complete code structures.

\section{Conclusion} \label{C}

This paper presented FLUC, a modular framework that integrates open-source LLMs with the ArduPilot autopilot to enable natural language-based UAV mission control. By translating high-level operator intent into executable code, FLUC lowers the entry barrier for autonomous UAV operations across domains such as wireless networking, surveillance, and environmental monitoring.

FLUC was evaluated using three open-source LLMs -- Qwen 2.5, Gemma 2, and LLaMA 3.2 -- across five representative scenarios and a case study involving energy-aware UAV positioning. Qwen 2.5 demonstrated the most consistent and accurate performance in code generation and structured reasoning. Gemma 2 achieved a favorable balance between output quality and computational efficiency, while LLaMA 3.2 offered faster responses but presented reduced reliability in more complex tasks.

Future work will extend FLUC with perception-driven awareness of UAV behaviour in response to generated code, as well as autonomous error recovery mechanisms to enhance adaptability and scalability in real-world deployments.



\bibliographystyle{IEEEtran}
\bibliography{refs}

\end{document}